\documentclass{article}
\usepackage{graphicx}
\usepackage[margin=1in]{geometry}
\usepackage{amsmath}
\usepackage{amssymb}
\usepackage{amsthm}
\usepackage{accents}
\usepackage[maxfloats=128]{morefloats}
\usepackage[breaklinks,hidelinks]{hyperref}

\newtheorem{theorem}{Theorem}[]

\newtheorem{remark1}[theorem]{Remark}

\title{Methods of interpreting error estimates\\
for grayscale image reconstructions}
\author{Aaron Defazio and Mark Tygert}

\begin{document}

\maketitle

\begin{abstract}
One representation of possible errors in a grayscale image reconstruction
is as another grayscale image estimating potentially worrisome differences
between the reconstruction and the actual ``ground-truth'' reality.
Visualizations and summary statistics can aid in the interpretation
of such a representation of error estimates.
Visualizations include suitable colorizations of the reconstruction,
as well as the obvious ``correction'' of the reconstruction
by subtracting off the error estimates.
The canonical summary statistic would be the root-mean-square
of the error estimates.
Numerical examples involving cranial magnetic-resonance imaging
clarify the relative merits of the various methods
in the context of compressed sensing.
Unfortunately, the colorizations appear likely to be too distracting
for actual clinical practice, and the root-mean-square gets swamped
by background noise in the error estimates.
Fortunately, straightforward displays of the error estimates
and of the ``corrected'' reconstruction are illuminating,
and the root-mean-square improves greatly after mild blurring
of the error estimates; the blurring is barely perceptible to the human eye
yet smooths away background noise that would otherwise overwhelm
the root-mean-square.
\end{abstract}

\section{Introduction}

Compressed sensing in imaging is a paradigm for accelerating the acquisition
of full images by taking fewer measurements than the number of degrees
of freedom being reconstructed. The measurements are thus ``undersampled''
relative to the usual information-theoretic requirements of sampling
at the Nyquist rate etc. Compressed sensing therefore risks introducing errors,
errors which very well may vary among different image acquisitions.
Recent work of~\cite{tygert-ward-zbontar} and others generates an error ``bar''
for each reconstructed image, in the form of another image that can be expected
to be representative of potential differences between the reconstruction
and the real ground-truth. The present paper considers user-friendly methods
for generating visualizations and automatic interpretations
of these error estimates,
appropriate for display to medical professionals (especially radiologists)
on data of cranial scans from magnetic-resonance imaging (MRI) machines.

After testing several natural visual displays,
we find that any nontrivial visualization is likely to be too distracting
for physicians, as some have expressed reservations about having to look
at any errors at all --- they would be much happier having a machine look
at the estimates and flag potentially serious errors for special consideration.
We might conclude that colorization is too distracting,
that the best visualizations are simple displays of the error estimates,
possibly supplemented with the error estimates
subtracted from the reconstructions (thus showing how the error estimates
can ``correct'' the reconstructions).
Most of the results of the present paper about visualization could be regarded
as negative, however natural and straightforward the colorizations may be.

For circumstances in which visualizing errors is overkill
(or unnecessarily bothersome),
we find that an almost simplistic automated interpretation
of the plots of errors --- reporting just the root-mean-square
of the denoised error estimates --- works remarkably well.
While background noise dominates the root-mean-square
of the initial, noisy error estimates,
even denoising that is almost imperceptible
can remove the obfuscatory background noise;
the root-mean-square can then focus on the remaining errors,
which are often relatively sparsely distributed.
When the root-mean-square of the denoised error estimates is large enough,
a clinician could look at the visualizations mentioned above
to fully understand the implications of the error estimates
(or rescan the patient using a less error-prone sampling pattern).

\section{Methods}

\subsection{Visualization in grayscale and in color}
\label{viz}

We include four kinds of plots displaying the full reconstructions
and errors:
\begin{enumerate}
\item ``Original'' is the original grayscale image.
\item ``Reconstruction'' is the reconstruction via compressed sensing.
Specifically, we use the configuration of~\cite{tygert-ward-zbontar};
for details (which are largely irrelevant for comparing the utility
of visualizations), please see the third section, ``Numerical examples,''
of~\cite{tygert-ward-zbontar}.
\item ``Error of Reconstruction'' displays the difference
between the original and reconstructed images, with black (or white)
corresponding to extreme errors, and middling grays
corresponding to the absence of errors.
\item ``Bootstrap'' displays the errors estimated via the bootstrap
of~\cite{tygert-ward-zbontar} (using the same 1000 iterations used
by~\cite{tygert-ward-zbontar}),
with black (or white) corresponding to extreme errors,
and middling grays corresponding to the absence of errors.
\end{enumerate}

We visualize the errors in reconstruction and the bootstrap estimates
using grayscale so that the phases of oscillatory artifacts are less apparent;
colorized errors look very different for damped sine versus cosine waves,
whereas the medical meaning of such waves is often similar.
Appendix~\ref{colors} displays the errors in color.

We consider four methods for visualizing the effects of errors
(as estimated via the bootstrap) simultaneously with displaying
the reconstruction, via manipulation of the hue-saturation-value color space
described, for example, by~\cite{scikit-image}:
\begin{enumerate}
\item ``Reconstruction - Bootstrap'' is literally the bootstrap error estimate
subtracted from the reconstruction, in some sense ``correcting''
or ``enhancing'' the reconstruction.
\item ``Errors Over a Threshold Overlaid'' identifies the pixels
in the bootstrap error estimate whose absolute values
are in the upper percentiles (the upper two percentiles
for horizontally retained sampling,
the upper one for radially retained sampling), then replaces those pixels
(retaining all other pixels unchanged) in the reconstruction with colors
corresponding to the values of the pixels in the bootstrap.
Specifically, the colors plotted are at the highest value possible
and fully saturated, with a hue ranging from cyan to magenta,
with blue in the middle (however, as we include only the upper percentiles,
only hues very close to cyan or to magenta actually get plotted).
This effectively marks the pixels corresponding to the largest estimated errors
with eye-popping colors, leaving the other pixels at their gray values
in the reconstruction.
\item ``Bootstrap-Saturated Reconstruction'' sets the saturation
of a pixel in the reconstruction to the corresponding absolute value
of the pixel in the bootstrap error estimate
(normalized by the greatest absolute value of any pixel in the bootstrap),
with a hue set to red or green depending on the sign of the pixel
in the bootstrap. The value of the pixel in the reconstruction stays the same.
Thus, a pixel gets colored more intensely red or more intensely green
when the absolute value of the pixel in the bootstrap is large,
but always with the value in hue-saturation-value remaining the same
as in the original reconstruction;
a pixel whose corresponding absolute value in the bootstrap
is relatively negligible stays unsaturated gray
at the value in the reconstruction.
\item ``Bootstrap-Interpolated Reconstruction'' leaves the value
of each pixel at its value in the reconstruction, and linearly interpolates
in the hue-saturation plane between green and magenta
based on the corresponding value of the pixel in the bootstrap error estimate
(normalized by the greatest absolute value of any pixel in the bootstrap).
Pure gray is in the middle of the line between green and magenta,
so that any pixel whose corresponding error estimate is zero
will appear unchanged, exactly as it was in the original reconstruction;
pixels whose corresponding error estimates are the largest
have the same value as in the reconstruction but get colored magenta,
while those whose corresponding error estimates are the most negative
have the same value as in the reconstruction but get colored green.
\end{enumerate}

\subsection{Summarization in a scalar}
\label{sum}

The square root of the sum of the squares of slightly denoised error
estimations summarizes in a single scalar the overall size of errors.
Even inconspicuous denoising can greatly improve the root-mean-square:
While the effect of blurring the bootstrap error estimates
with a normalized Gaussian convolutional kernel of standard deviation one pixel
is almost imperceptible to the human eye (or at least preserves
the semantically meaningful structures in the images),
the blur helps remove the background of noise that can otherwise dominate
the root-mean-square of the error estimates.
The blur largely preserves significant edges and textured areas, yet
can eliminate much of the perceptually immaterial zero-mean background noise.
Whereas background noise can overwhelm the root-mean-square
of the initial, noisy bootstrap, the root-mean-square
of the slightly blurred bootstrap captures the magnitude
of the important features in the error estimates.

\section{Results}
\label{results}

Our data comes from~\cite{mri2}, \cite{mri1}, \cite{mri3}, \cite{mri4}.
Specifically, we consider two cross-sectional slices through the head
of a patient in an MRI scanner:
the lower slice is the third of twenty from~\cite{tygert-ward-zbontar}, while
the upper slice is the tenth of twenty from~\cite{tygert-ward-zbontar}.
Compressed sensing reconstructs a cross-sectional image
given only a subset of the usual measurements of values of the two-dimensional
Fourier transform of the original, ``ground-truth'' cross-section.
We consider the radially retained and horizontally retained subsets
of~\cite{tygert-ward-zbontar}, which yield the error estimates displayed
in the figures below
(the third section, ``Numerical examples,'' of~\cite{tygert-ward-zbontar}
details the schemes for sampling, but these details are irrelevant
for assessment of visualizations).
We used the Python package (``fbooja'') of~\cite{tygert-ward-zbontar};
the Gaussian blur from Subsection~\ref{sum} leverages
skimage.filters.gaussian from scikit-image of~\cite{scikit-image}.

Figures~\ref{first_viz}--\ref{last_viz} display the visualizations
from Subsection~\ref{viz}.
Figures~\ref{blurredh} and~\ref{blurredr} depict the effects
of the Gaussian blur from Subsection~\ref{sum}.
Table~\ref{blurring} reports how drastically
such a nearly imperceptible blur changes
the square roots of the sums of the squares of the error estimates.
Background noise clearly overwhelms the root-mean-square without any denoising
of the error estimates --- the root-mean-square decreases dramatically
even with just the mild denoising of blurring
with a normalized Gaussian convolutional kernel
whose standard deviation is one pixel, as in Table~\ref{blurring}
and Figures~\ref{blurredh} and~\ref{blurredr}.
Tables~\ref{blurs3} and~\ref{blurs10} report how blurring with wider Gaussians
affects the root-mean-square; of course, wider Gaussian blurs
are much more conspicuous and risk washing out important coherent features
of the error estimates, while the last column of Table~\ref{blurs10} shows
that denoising with wider Gaussian blurs brings diminishing returns.
The width used in Table~\ref{blurring} and Figures~\ref{blurredh}
and~\ref{blurredr} --- only one pixel --- may be safest.
Appendix~\ref{blurredover} displays the grayscale reconstructions overlaid
with the blurred bootstraps (blurring with a Gaussian whose standard deviation
is one pixel), thresholded and colorized as in Subsection~\ref{viz}.

\section{Discussion and Conclusion}

Broadly speaking, the bootstrap-saturated reconstructions
and bootstrap-interpolated reconstructions look similar,
even though the details of their constructions differ.
Both the bootstrap-saturated reconstruction and the bootstrap-interpolated
reconstruction highlight errors more starkly on pixels
for which the reconstruction is bright; dark green, dark red, and dark magenta
(that is, with a relatively low value in hue-saturation-value) simply do not
jump out visually, even if the green, red, or magenta are fully saturated.
That said, retaining the value of the pixel in the reconstruction makes
the colorization of the bootstrap-saturated reconstruction
and the bootstrap-interpolated reconstruction far less distracting
than in errors over a threshold overlaid, with much higher fidelity
to the form of the grayscale reconstruction in the colored regions.
Of course, the errors over a threshold overlaid do not alter
the grayscale reconstruction at all when the errors are within the threshold,
so the fidelity to the grayscale reconstruction is perfect
in those areas of the images with overlaid errors where the error estimates
do not go beyond the threshold.

Thus, none of the colorizations is uniformly superior to the others,
and all may be too distracting for actual clinical practice.
Alternatives include direct display of the bootstrap error estimates,
possibly complemented by the bootstrap subtracted from the reconstruction
(to illustrate the effects of ``correcting'' the reconstruction
with the error estimates), which are readily interpretable
and minimally distracting.

The bootstrap subtracted from the reconstruction tends to sharpen
the reconstruction and to add back some features such as lines or textures
that the reconstruction obscured. However, this reconstruction that is
``corrected'' with the bootstrap estimations may contain artifacts
not present in the original image --- the error estimates
tend to be conservative, possibly suspecting errors in some regions
where in fact the reconstruction is accurate. The ``corrected'' reconstruction
(that is, the bootstrap subtracted from the reconstruction)
can be illuminating, but only as a complement
to plotting the bootstrap error estimates on their own, too.

A sensible protocol could be to check if the root-mean-square
of the blurred bootstrap is large enough to merit further investigation,
investigating further by looking at the full bootstrap image
together with the reconstruction ``corrected''
by subtracting off the bootstrap error estimates (or colorizations).

\begin{table}
\caption{square roots of the sums of the squares of the error estimates}
\label{blurring}
\begin{centering}

\vspace{.125in}

\begin{tabular}{llll}
    Sampling & Slice & Bootstrap & Blurred Bootstrap \\\hline
horizontally & lower &      12.9 &              6.25 \\
horizontally & upper &      13.8 &              7.34 \\
    radially & lower &      17.5 &              10.5 \\
    radially & upper &      18.0 &              11.6
\end{tabular}

\end{centering}
\end{table}

\begin{table}

\vspace{.125in}

\caption{square roots of the sums of the squares of the error estimates
for the lower slice blurred against a Gaussian convolutional kernel
of the specified standard deviation (the standard deviation is in pixels),
for sampling retained horizontally or radially}
\label{blurs3}
\begin{centering}

\vspace{.125in}

\begin{tabular}{lll}
Std.\ Dev. & Horizontally & Radially \\\hline
       0.0 &         12.9 &     17.5 \\
       0.5 &         9.94 &     14.6 \\
       1.0 &         6.25 &     10.5 \\
       1.5 &         4.38 &     8.06 \\
       2.0 &         3.03 &     6.34 \\
       2.5 &         2.04 &     5.06 \\
       3.0 &         1.33 &     4.09 \\
       3.5 &         .847 &     3.34 \\
       4.0 &         .535 &     2.75
\end{tabular}

\end{centering}
\end{table}

\begin{table}

\vspace{.125in}

\caption{square roots of the sums of the squares of the error estimates
for the upper slice blurred against a Gaussian convolutional kernel
of the specified standard deviation (the standard deviation is in pixels),
for sampling retained horizontally or radially}
\label{blurs10}
\begin{centering}

\vspace{.125in}

\begin{tabular}{lll}
Std.\ Dev. & Horizontally & Radially \\\hline
       0.0 &         13.8 &     18.0 \\
       0.5 &         10.9 &     15.3 \\
       1.0 &         7.34 &     11.6 \\
       1.5 &         5.35 &     9.50 \\
       2.0 &         3.82 &     7.97 \\
       2.5 &         2.63 &     6.79 \\
       3.0 &         1.75 &     5.87 \\
       3.5 &         1.14 &     5.13 \\
       4.0 &         .745 &     4.54
\end{tabular}

\end{centering}
\end{table}

\newlength{\vertsep}
\setlength{\vertsep}{.25in}
\newlength{\imsize}
\setlength{\imsize}{.47\textwidth}

\begin{figure}
\begin{centering}

\parbox{\imsize}{\includegraphics[width=\imsize]{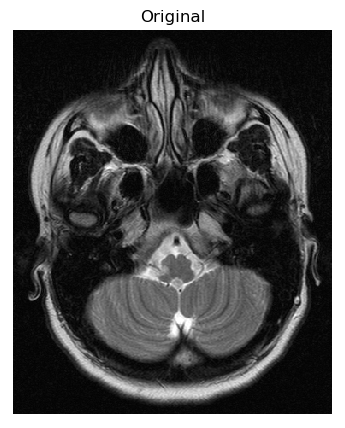}}
\hfill
\parbox{\imsize}{\includegraphics[width=\imsize]{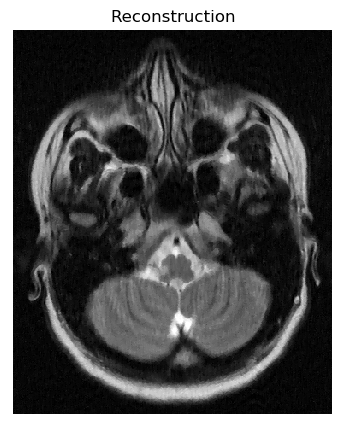}}

\vspace{\vertsep}

\parbox{\imsize}{\includegraphics[width=\imsize]{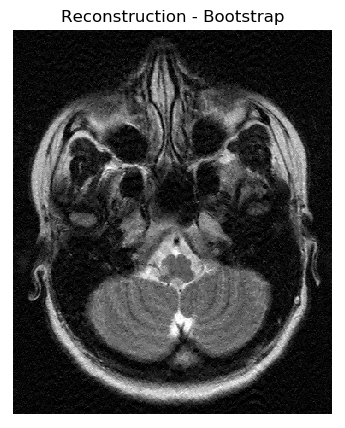}}
\hfill
\parbox{\imsize}{\includegraphics[width=\imsize]{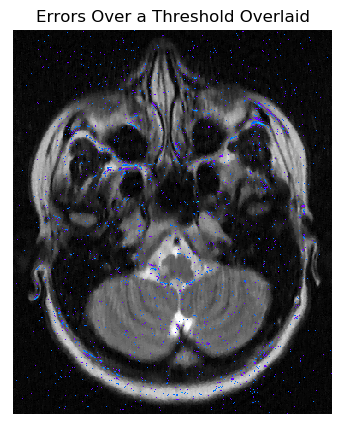}}

\end{centering}
\caption{horizontally retained sampling --- lower slice (a)}
\label{first_viz}
\end{figure}

\begin{figure}
\begin{centering}

\parbox{\imsize}{\includegraphics[width=\imsize]{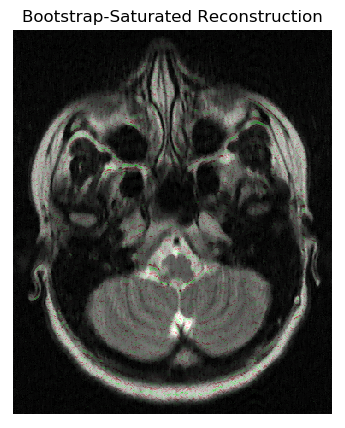}}
\hfill
\parbox{\imsize}{\includegraphics[width=\imsize]{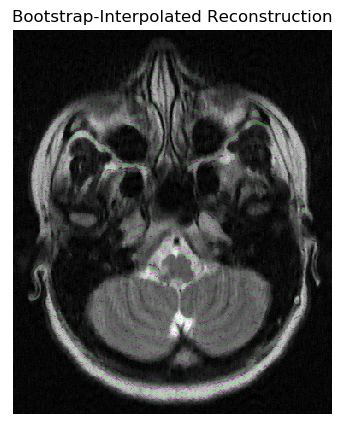}}

\vspace{\vertsep}

\parbox{\imsize}{\includegraphics[width=\imsize]{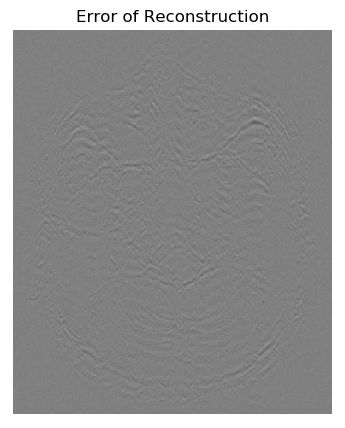}}
\hfill
\parbox{\imsize}{\includegraphics[width=\imsize]{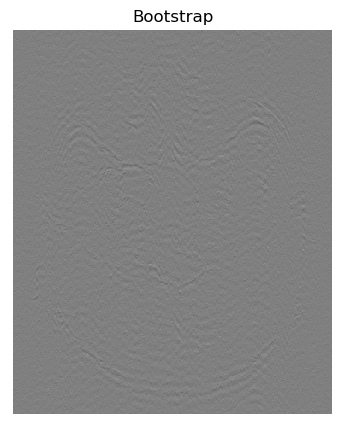}}

\end{centering}
\caption{horizontally retained sampling --- lower slice (b)}
\end{figure}

\begin{figure}
\begin{centering}

\parbox{\imsize}{\includegraphics[width=\imsize]{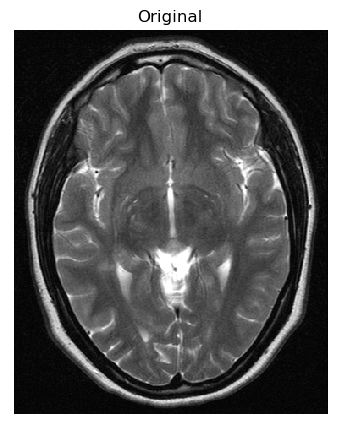}}
\hfill
\parbox{\imsize}{\includegraphics[width=\imsize]{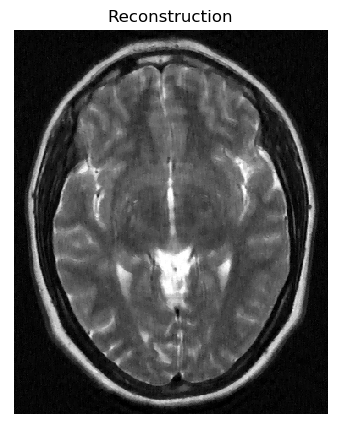}}

\vspace{\vertsep}

\parbox{\imsize}{\includegraphics[width=\imsize]{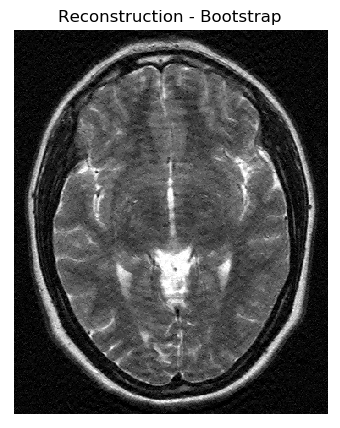}}
\hfill
\parbox{\imsize}{\includegraphics[width=\imsize]{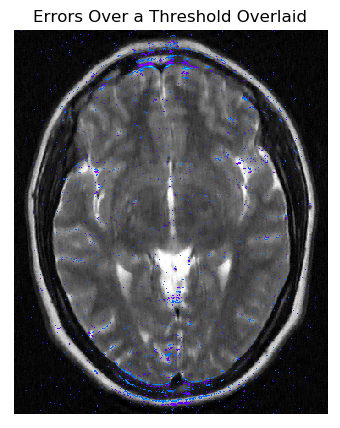}}

\end{centering}
\caption{horizontally retained sampling --- upper slice (a)}
\end{figure}

\begin{figure}
\begin{centering}

\parbox{\imsize}{\includegraphics[width=\imsize]{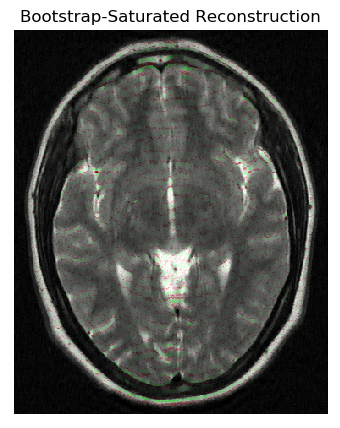}}
\hfill
\parbox{\imsize}{\includegraphics[width=\imsize]{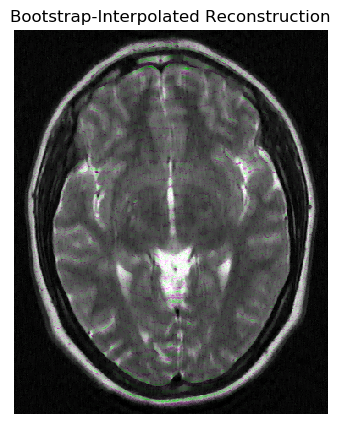}}

\vspace{\vertsep}

\parbox{\imsize}{\includegraphics[width=\imsize]{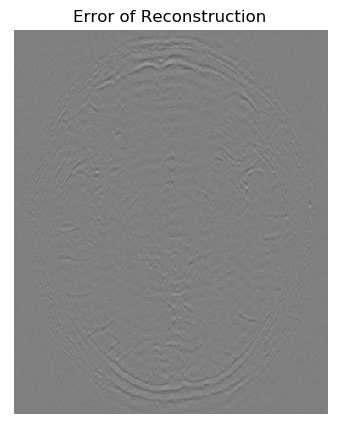}}
\hfill
\parbox{\imsize}{\includegraphics[width=\imsize]{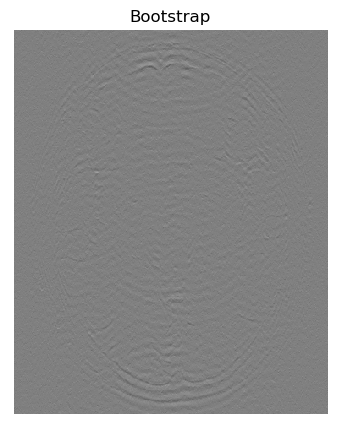}}

\end{centering}
\caption{horizontally retained sampling --- upper slice (b)}
\end{figure}

\begin{figure}
\begin{centering}

\parbox{\imsize}{\includegraphics[width=\imsize]{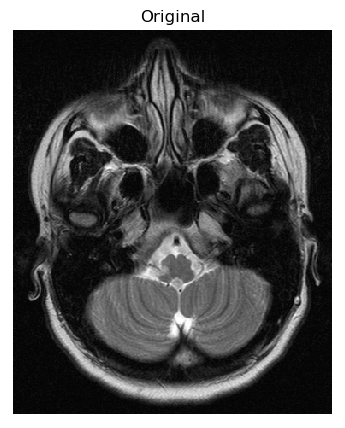}}
\hfill
\parbox{\imsize}{\includegraphics[width=\imsize]{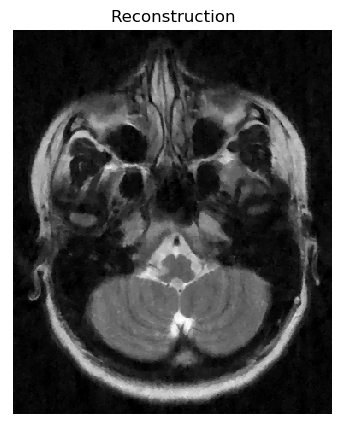}}

\vspace{\vertsep}

\parbox{\imsize}{\includegraphics[width=\imsize]{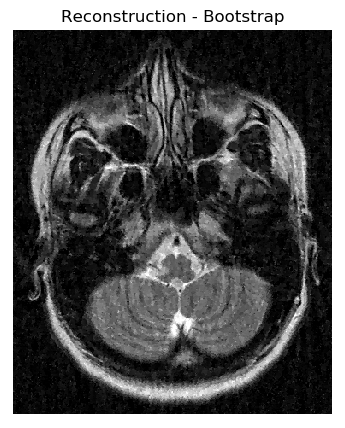}}
\hfill
\parbox{\imsize}{\includegraphics[width=\imsize]{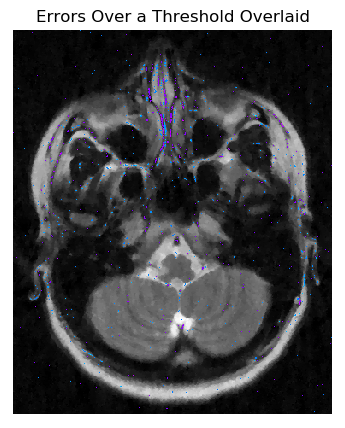}}

\end{centering}
\caption{radially retained sampling --- lower slice (a)}
\end{figure}

\begin{figure}
\begin{centering}

\parbox{\imsize}{\includegraphics[width=\imsize]{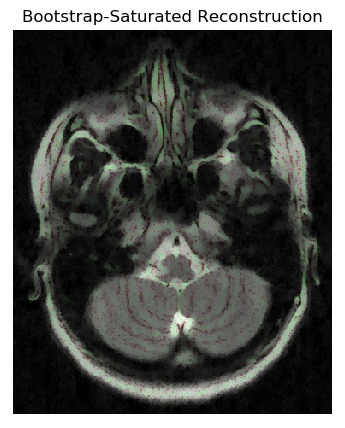}}
\hfill
\parbox{\imsize}{\includegraphics[width=\imsize]{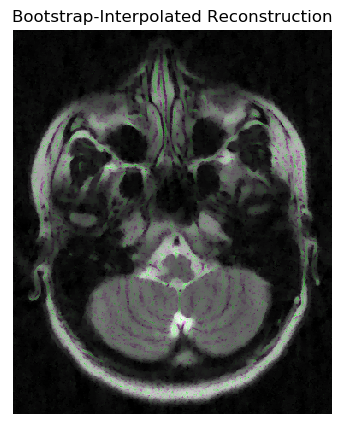}}

\vspace{\vertsep}

\parbox{\imsize}{\includegraphics[width=\imsize]{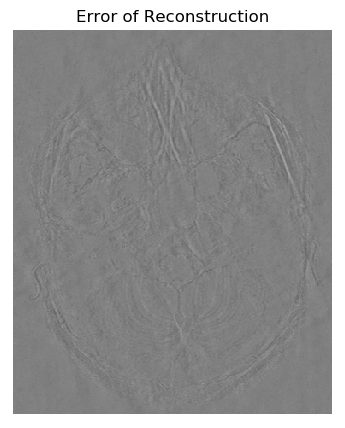}}
\hfill
\parbox{\imsize}{\includegraphics[width=\imsize]{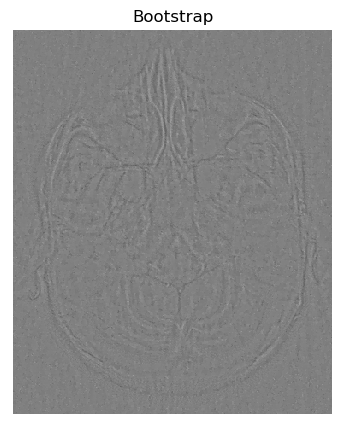}}

\end{centering}
\caption{radially retained sampling --- lower slice (b)}
\end{figure}

\begin{figure}
\begin{centering}

\parbox{\imsize}{\includegraphics[width=\imsize]{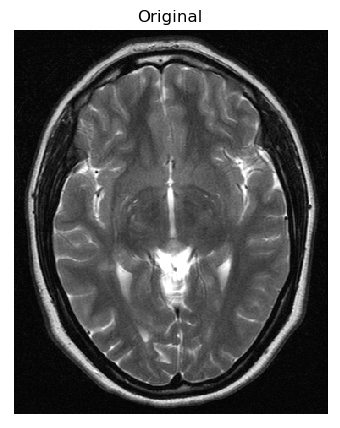}}
\hfill
\parbox{\imsize}{\includegraphics[width=\imsize]{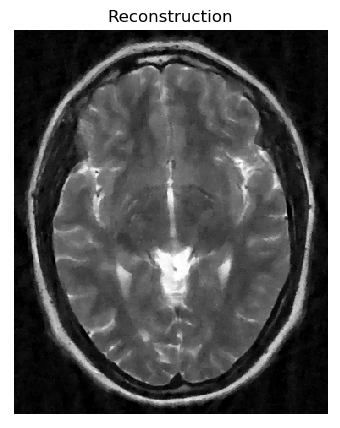}}

\vspace{\vertsep}

\parbox{\imsize}{\includegraphics[width=\imsize]{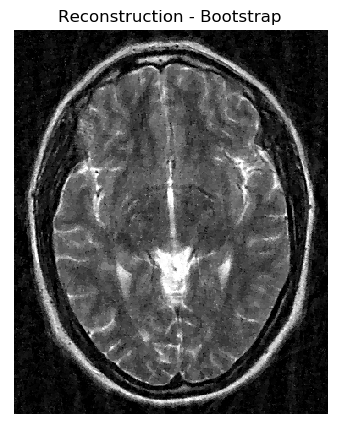}}
\hfill
\parbox{\imsize}{\includegraphics[width=\imsize]{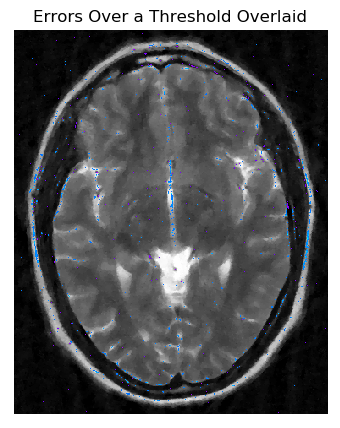}}

\end{centering}
\caption{radially retained sampling --- upper slice (a)}
\end{figure}

\begin{figure}
\begin{centering}

\parbox{\imsize}{\includegraphics[width=\imsize]{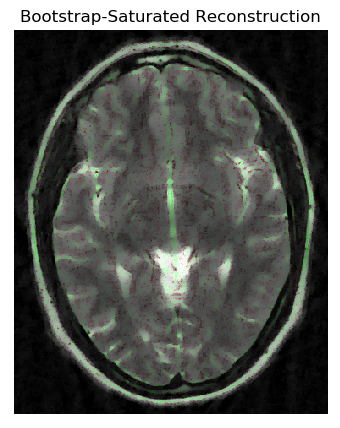}}
\hfill
\parbox{\imsize}{\includegraphics[width=\imsize]{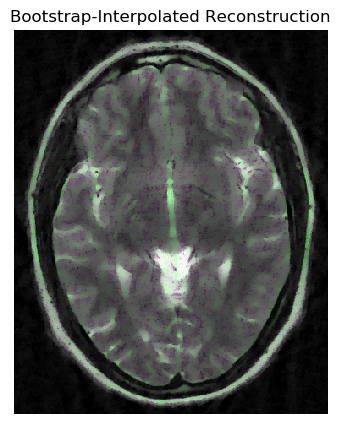}}

\vspace{\vertsep}

\parbox{\imsize}{\includegraphics[width=\imsize]{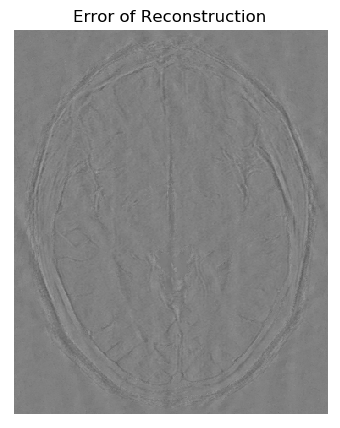}}
\hfill
\parbox{\imsize}{\includegraphics[width=\imsize]{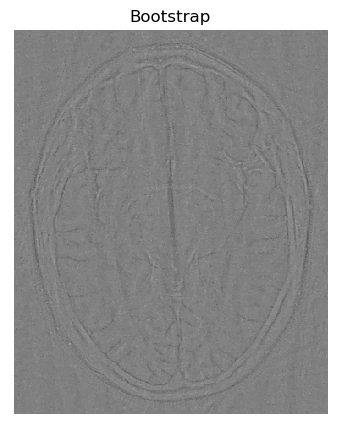}}

\end{centering}
\caption{radially retained sampling --- upper slice (b)}
\label{last_viz}
\end{figure}

\begin{figure}
\begin{centering}

\parbox{\imsize}{\includegraphics[width=\imsize]{bootstrap10h}}
\hfill
\parbox{\imsize}{\includegraphics[width=\imsize]{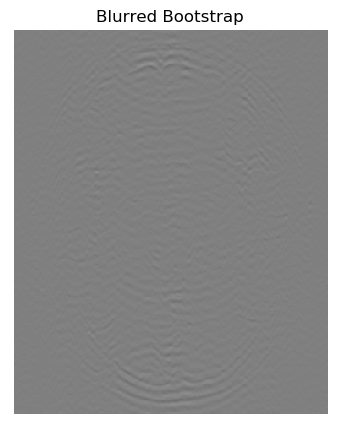}}

\vspace{\vertsep}

\parbox{\imsize}{\includegraphics[width=\imsize]{bootstrap3h}}
\hfill
\parbox{\imsize}{\includegraphics[width=\imsize]{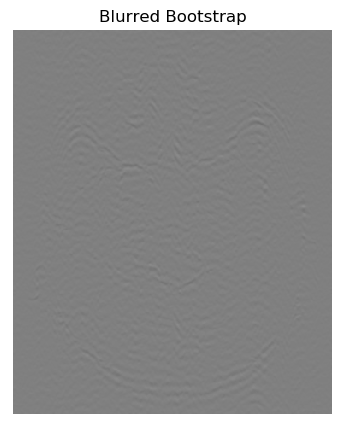}}

\end{centering}
\caption{horizontally retained sampling
--- upper plots display the upper slice; lower plots display the lower}
\label{blurredh}
\end{figure}

\begin{figure}
\begin{centering}

\parbox{\imsize}{\includegraphics[width=\imsize]{bootstrap10r}}
\hfill
\parbox{\imsize}{\includegraphics[width=\imsize]{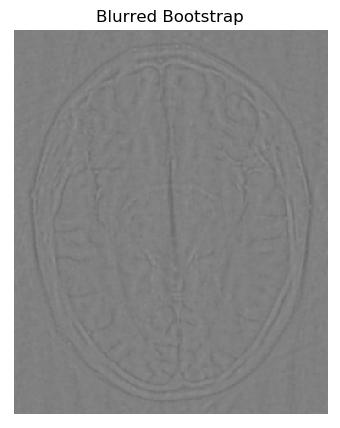}}

\vspace{\vertsep}

\parbox{\imsize}{\includegraphics[width=\imsize]{bootstrap3r}}
\hfill
\parbox{\imsize}{\includegraphics[width=\imsize]{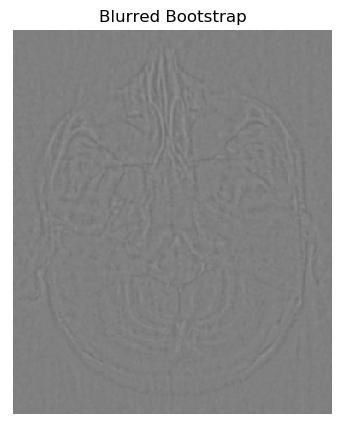}}

\end{centering}
\caption{radially retained sampling
--- upper plots display the upper slice; lower plots display the lower}
\label{blurredr}
\end{figure}

\clearpage

\appendix
\section{Bootstraps and Errors in Color}
\label{colors}

For reference, this appendix displays the errors of reconstruction
and bootstrap estimates in color, with blue for negative errors,
red for positive errors, and white for the absence of any error
(light blue and light red indicate less extreme errors
than pure blue or pure red).
The labeling conventions (``lower,'' ``upper,'' etc.)
conform to those introduced in Section~\ref{results}.

\begin{figure}
\begin{centering}

\parbox{\imsize}{\includegraphics[width=\imsize]{original3h}}
\hfill
\parbox{\imsize}{\includegraphics[width=\imsize]{recon3h}}

\vspace{\vertsep}

\parbox{\imsize}{\includegraphics[width=\imsize]{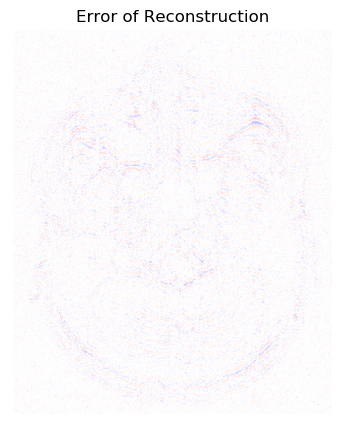}}
\hfill
\parbox{\imsize}{\includegraphics[width=\imsize]{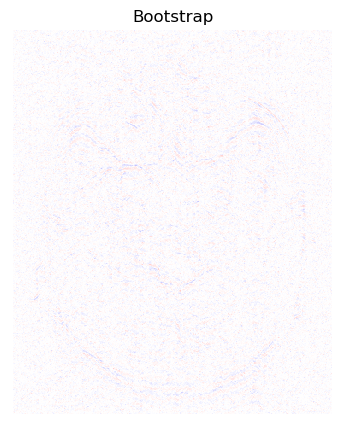}}

\end{centering}
\caption{horizontally retained sampling --- lower slice (c)}
\end{figure}

\begin{figure}
\begin{centering}

\parbox{\imsize}{\includegraphics[width=\imsize]{original10h}}
\hfill
\parbox{\imsize}{\includegraphics[width=\imsize]{recon10h}}

\vspace{\vertsep}

\parbox{\imsize}{\includegraphics[width=\imsize]{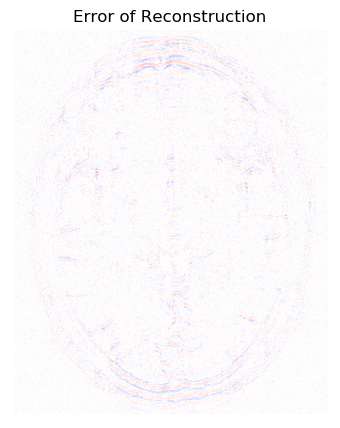}}
\hfill
\parbox{\imsize}{\includegraphics[width=\imsize]{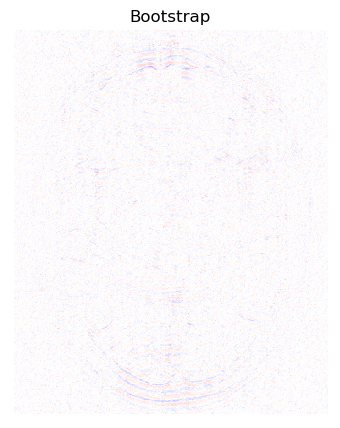}}

\end{centering}
\caption{horizontally retained sampling --- upper slice (c)}
\end{figure}

\begin{figure}
\begin{centering}

\parbox{\imsize}{\includegraphics[width=\imsize]{original3r}}
\hfill
\parbox{\imsize}{\includegraphics[width=\imsize]{recon3r}}

\vspace{\vertsep}

\parbox{\imsize}{\includegraphics[width=\imsize]{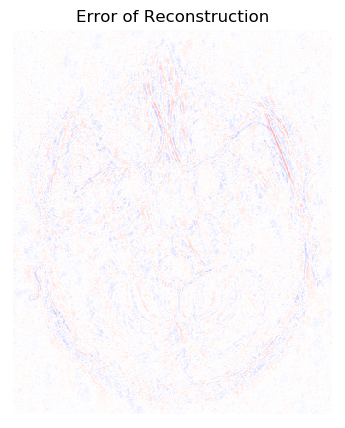}}
\hfill
\parbox{\imsize}{\includegraphics[width=\imsize]{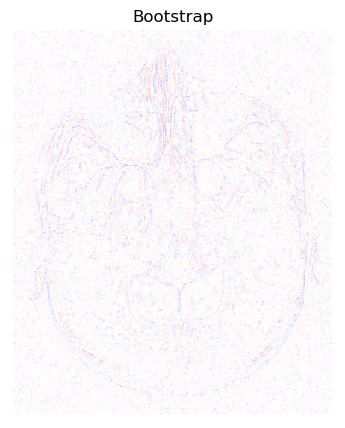}}

\end{centering}
\caption{radially retained sampling --- lower slice (c)}
\end{figure}

\begin{figure}
\begin{centering}

\parbox{\imsize}{\includegraphics[width=\imsize]{original10r}}
\hfill
\parbox{\imsize}{\includegraphics[width=\imsize]{recon10r}}

\vspace{\vertsep}

\parbox{\imsize}{\includegraphics[width=\imsize]{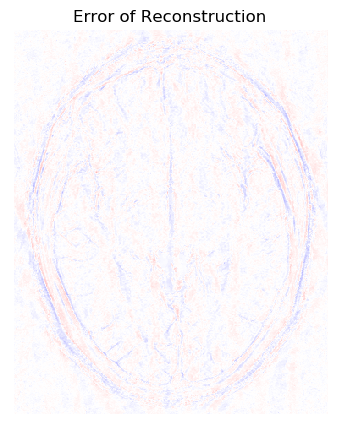}}
\hfill
\parbox{\imsize}{\includegraphics[width=\imsize]{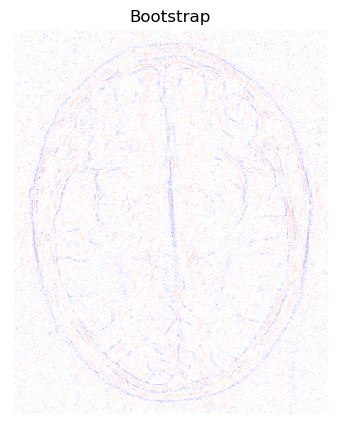}}

\end{centering}
\caption{radially retained sampling --- upper slice (c)}
\end{figure}

\clearpage

\section{Blurred Errors Over a Threshold Overlaid}
\label{blurredover}

For reference, this appendix displays the same errors over a threshold overlaid
over the reconstruction as in Subsection~\ref{viz},
together with the blurred errors over a threshold overlaid
over the reconstruction (blurring with a Gaussian convolutional kernel
whose standard deviation is one pixel, as in Subsection~\ref{sum}).
The labeling conventions (``lower,'' ``upper,'' etc.)
conform to those introduced in Section~\ref{results}.
The blurred errors certainly introduce less distracting noise
than without blurring, yet the colors still appear really distracting.

\begin{figure}
\begin{centering}

\parbox{\imsize}{\includegraphics[width=.945\imsize]{overlaid10h}}
\hfill
\parbox{\imsize}{\includegraphics[width=\imsize]{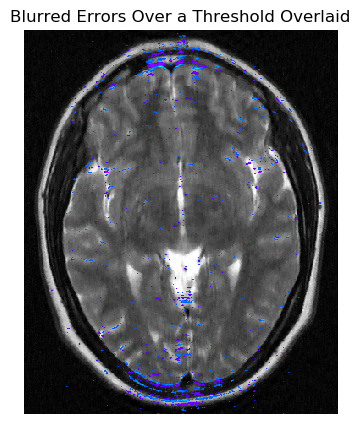}}

\vspace{\vertsep}

\parbox{\imsize}{\includegraphics[width=.96\imsize]{overlaid3h}}
\hfill
\parbox{\imsize}{\includegraphics[width=\imsize]{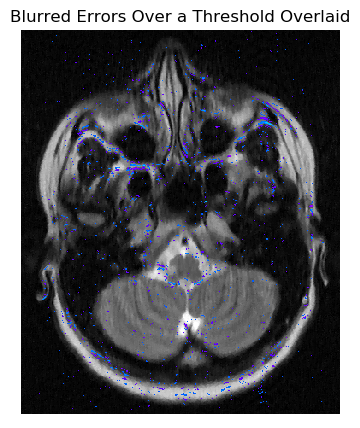}}

\end{centering}
\caption{horizontally retained sampling (b)
--- upper plots show the upper slice; lower plots show the lower}
\end{figure}

\begin{figure}
\begin{centering}

\parbox{\imsize}{\includegraphics[width=.945\imsize]{overlaid10r}}
\hfill
\parbox{\imsize}{\includegraphics[width=\imsize]{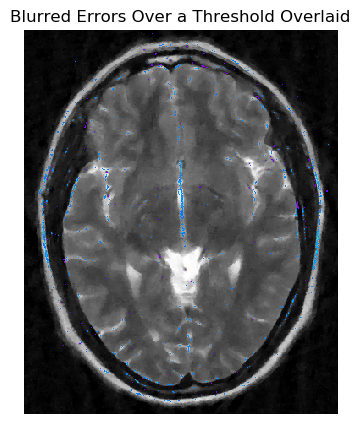}}

\vspace{\vertsep}

\parbox{\imsize}{\includegraphics[width=.96\imsize]{overlaid3r}}
\hfill
\parbox{\imsize}{\includegraphics[width=\imsize]{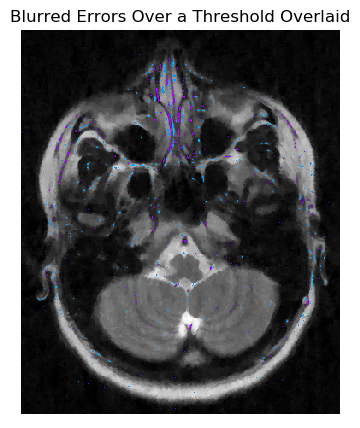}}

\end{centering}
\caption{radially retained sampling (b)
--- upper plots display the upper slice; lower plots display the lower}
\end{figure}

\clearpage

\bibliography{viz}
\bibliographystyle{apalike}

\end{document}